\documentclass[10pt,twocolumn]{article}
\usepackage[a4paper,margin=0.6in]{geometry}
\usepackage{cite}
\usepackage{amsmath,amssymb,amsfonts}
\usepackage{graphicx}
\usepackage{textcomp}
\usepackage{array}
\usepackage{url}
\usepackage{verbatim}
\usepackage{listings}
\usepackage{balance}
\usepackage{caption}
\usepackage{subcaption}
\usepackage{xcolor}
\usepackage{gensymb}
\usepackage{afterpage}
\usepackage{dblfloatfix}
\usepackage{algorithm}
\usepackage{algpseudocode}

\algtext*{EndIf} % Remove "End if"
\algtext*{EndFor} % Remove "End for"
\algtext*{EndProcedure} % Remove "End procedure"
\def\BibTeX{{\rm B\kern-.05em{\sc i\kern-.025em b}\kern-.08em
    T\kern-.1667em\lower.7ex\hbox{E}\kern-.125emX}}
    
\begin{document}

\newcommand{\arxivnotice}{%
\copyright~2025 IEEE. This paper has been accepted for presentation at the Design, Automation and Test in Europe Conference (DATE) 2025. The official version will appear in the DATE 2025 proceedings.
}

\title{\textbf{An ASIC Emulated Oscillator Ising/Potts Machine Solving Combinatorial Optimization Problems}}

\author{Yilmaz Ege Gonul, Baris Taskin \\ \textit{Drexel University, Philadelphia, PA, USA}\\ Email: \{yeg26,bt62\}@drexel.edu}

\date{} 

\maketitle

\footnotetext{
© 2026 IEEE. This paper has been accepted for presentation at the IEEE International Symposium on Circuits and Systems
 2026. The official version will appear in the ISCAS 2026 proceedings.
}

\begin{abstract}
Oscillator-based Ising/Potts machines (OIMs/OPMs) are promising hardware accelerators for NP-hard combinatorial optimization problems using coupled oscillator synchronization dynamics. Analog OIMs/OPMs offer speed advantages but have limited coupling resolution, process variation susceptibility, and scalability issues, while digital GPU/CPU emulations provide flexibility but suffer from irregular memory access patterns and energy inefficiency. This work presents a custom ASIC architecture that digitally emulates OIM/OPM dynamics using simplified fixed-point Kuramoto model equations. The scalable design features processing elements with direct interconnections, eliminating shared memory bottleneck while maintaining digital programmability and precision. A 20×20 processing element array with king's graph connectivity is prototyped and evaluated via post-layout simulations on unweighted/weighted max-cut and graph coloring problems, achieving 97-100\% maximum accuracy with significant speed and energy improvements over general-purpose platforms, demonstrating the viability of algorithmically codesigned ASICs.
\end{abstract}


\section{Introduction}

Combinatorial optimization problems (COPs) represent a critical class of computational challenges with applications spanning various fields. The Ising model~\cite{ising_model} provides a framework for encoding these problems as energy minimization tasks, where the objective is to minimize a Hamiltonian $H \propto -\sum_{i<j} J_{ij}s_is_j$ with binary spins $s_i \in \{-1, +1\}$ and problem-specific coupling strengths $J_{ij}$. The Potts model~\cite{potts_model} extends these dynamics to multi-valued spins. Many NP-hard problems, including max-cut, graph coloring, and satisfiability, can be mapped to these forms~\cite{many_ising}. Implementations of Ising/Potts models on conventional Von Neumann architectures struggle with large instances, motivating specialized hardware accelerators~\cite{COP_exponential}.

Ising/Potts machines are specialized hardware designed to find low-energy configurations by exploiting physical dynamics. Various implementations have been proposed, including quantum annealers~\cite{dwave,dwave2}, optical systems~\cite{CIM,coherent_potts,potts_nature}, and oscillator-based implementations~\cite{wang_oim}, some leveraging CMOS technology~\cite{prob_fabric, chris_kim_nat,date_potts,iccad_potts,rtwo_ising}. Oscillator-based Ising/Potts machines (OIMs/OPMs) utilize coupled oscillator networks whose Kuramoto-model phase dynamics~\cite{wang_oim} naturally synchronize in patterns corresponding to low-energy states, offering inherent parallelism through simultaneous evolution of all oscillator states.

Despite their promise, existing OIM/OPM implementations face significant limitations. Analog OIMs/OPMs made of LC oscillators~\cite{wang_oim} or ring oscillators~\cite{prob_fabric,iccad_potts} can achieve fast dynamics but suffer from process variations, limited in programmability, scalability and connectivity due to fanout limitations and the lack of time-sliceability. Digital emulations on GPUs~\cite{gpu_gls} and CPUs address programmability and resolution concerns, but the former is constrained by irregular memory access patterns and global memory bottlenecks, whereas the latter by limitations in parallelism. 
%Both platforms are also inefficient for COP workloads in terms of resource allocation, latency, and energy consumption.  

This work presents a custom ASIC that combines digital precision and programmability with the performance of specialized hardware. The algorithm co-design digitally emulates simplified Kuramoto dynamics using fixed-point arithmetic, reducing coupling and synchronization functions to efficient arithmetic and logic operations. An array of interconnected processing elements (PEs) exchanges phase information directly, faster than the phase exchanges through global memory in conventional platforms. Building upon a previous similar work~\cite{emulation_oim}, the proposed scalable architecture is prototyped as a 20×20 array of PEs arranged in a king's graph topology. The architecture is extended to represent the Potts model, capable of mapping larger and multivariable COPs while reducing the hardware cost to 8-bit representation. 
Post-layout performance is evaluated on custom weighted and unweighted max-cut problems in the OIM mode, and a custom graph coloring problem in the OPM mode. The results demonstrate max solution accuracy in the range of 97\% to 100\% with significant improvements in speed and energy efficiency compared to existing approaches.

\section{Background}

This section introduces the mathematical foundations of oscillator-based Ising/Potts machines and the algorithmic simplifications that enable efficient hardware implementation.

\subsection{Oscillator Ising/Potts Machines and the Kuramoto Model}

OIMs/OPMs~\cite{wang_oim,seal_potts} leverage the synchronization dynamics of coupled oscillators to solve COPs. The behavior of such systems can be modeled using the generalized Kuramoto model~\cite{high_quality_oim,wang_oim2}, which describes the phase evolution of $n$ coupled oscillators while ensuring convergence to discrete phase states. The generalized Kuramoto equation can be expressed using an arbitrary coupling function $F_c$ operating on the phase differences, and an arbitrary synchronization function $F_s$ forcing the phases into discrete states:

\begin{equation}
\frac{d\phi_i}{dt} = -\sum_{j=1}^{n} J_{ij} \cdot F_c(\phi_i - \phi_j) - F_s(\phi_i)
\label{eq:eq1}
\end{equation}
where $\phi_i \in [0, 1)$ is the normalized phase of oscillator $i$, $J_{ij}$ is the coupling coefficient from the Ising problem. 

% $F_c(\psi)$ is a coupling function that determines how phase differences $\psi = \phi_i - \phi_j$ influence oscillator interactions, and $F_s(\phi)$ is a synchronization function that creates discrete stable states. 

In the original OIM/OPM formulations~\cite{wang_oim,seal_potts}, sinusoidal functions are used: $F_c(\psi) = \sin(2\pi\psi)$ for coupling where $\psi$ is the phase difference $(\phi_i - \phi_j)$ , and $F_s(\phi_i) = \sin(2\pi N\phi_i)$ for phase binarization/discretization with sub-harmonic injection locking (SHIL). Parameter $N$ determines the number of stable phase states. For Ising problems, $N=2$ creates two stable states at $\phi = 0$ and $\phi = 0.5$, corresponding to spins $s = +1$ and $s = -1$. Higher $N$ unlocks the Potts model~\cite{potts_model}, allowing multivalued spins to be represented via a single oscillator.

The coupling function $F_c$ creates phase-dependent interactions where oscillators tend to synchronize (for positive $J_{ij}$) or anti-synchronize (for negative $J_{ij}$). The synchronization function $F_s(\phi_i)$ introduces $N$ equally spaced potential wells in the phase space, drawing oscillators toward discrete fixed points while the coupling terms determine which wells different oscillators settle into based on mutual interactions.

% The oscillator-based physical model offers fundamental advantages as an Ising/Potts model realization. 

Simulating coupled oscillator dynamics inherently provides energy minimization as the system naturally evolves toward low-energy stable states without requiring explicit energy evaluation. The framework extends naturally to multi-valued Potts problems through parametric changes in the synchronization function $F_s$, enabling the same architecture to handle both binary and multivalued problems with diverse COP mappings defined by the coupling matrix $J$.

\subsection{Simplified Fixed-Point Implementation}

For efficient digital implementation, simplified piecewise constant functions are employed that approximate the behavior of sinusoidal $F_c$ and $F_s$ functions for Ising model~\cite{emulation_oim} requiring simpler arithmetic and logic operations. A simple coupling function $F_c$ can be defined as:

\begin{equation}
\label{eq:eq2}
F_c(\psi) = \begin{cases}
+1 & \text{if } (\psi \bmod 1) < 0.5 \\
-1 & \text{if } (\psi \bmod 1) \geq 0.5
\end{cases}
\end{equation}
where  $\psi$ is the phase difference ($\phi_i - \phi_j$).
A simple synchronization function $F_s^{(N)}$ that creates $N$ equally-spaced stable phase states is defined. For Ising problems ($N=2$):

\begin{equation}
\label{eq:eq3}
F_s^{(2)}(\phi) = \begin{cases}
-1 & \text{if } (\phi \bmod 0.5) < 0.25 \\
+1 & \text{if } (\phi \bmod 0.5) \geq 0.25
\end{cases}
\end{equation}

The proposed simplified $F_s$ extends to the 3-state Potts model, enabling a 3-state ($N=3$) representation: 
\begin{equation}
\label{eq:eq4}
\small
F_s^{(3)}(\phi) = \begin{cases}
-1 & \text{if } \phi \in [0, 1/6) \cup [1/3, 1/2) \cup [2/3, 5/6) \\
+1 & \text{if } \phi \in [1/6, 1/3) \cup [1/2, 2/3) \cup [5/6, 1)
\end{cases}
\end{equation}

\begin{figure}[t]
    \centering
    \includegraphics[width=\columnwidth]{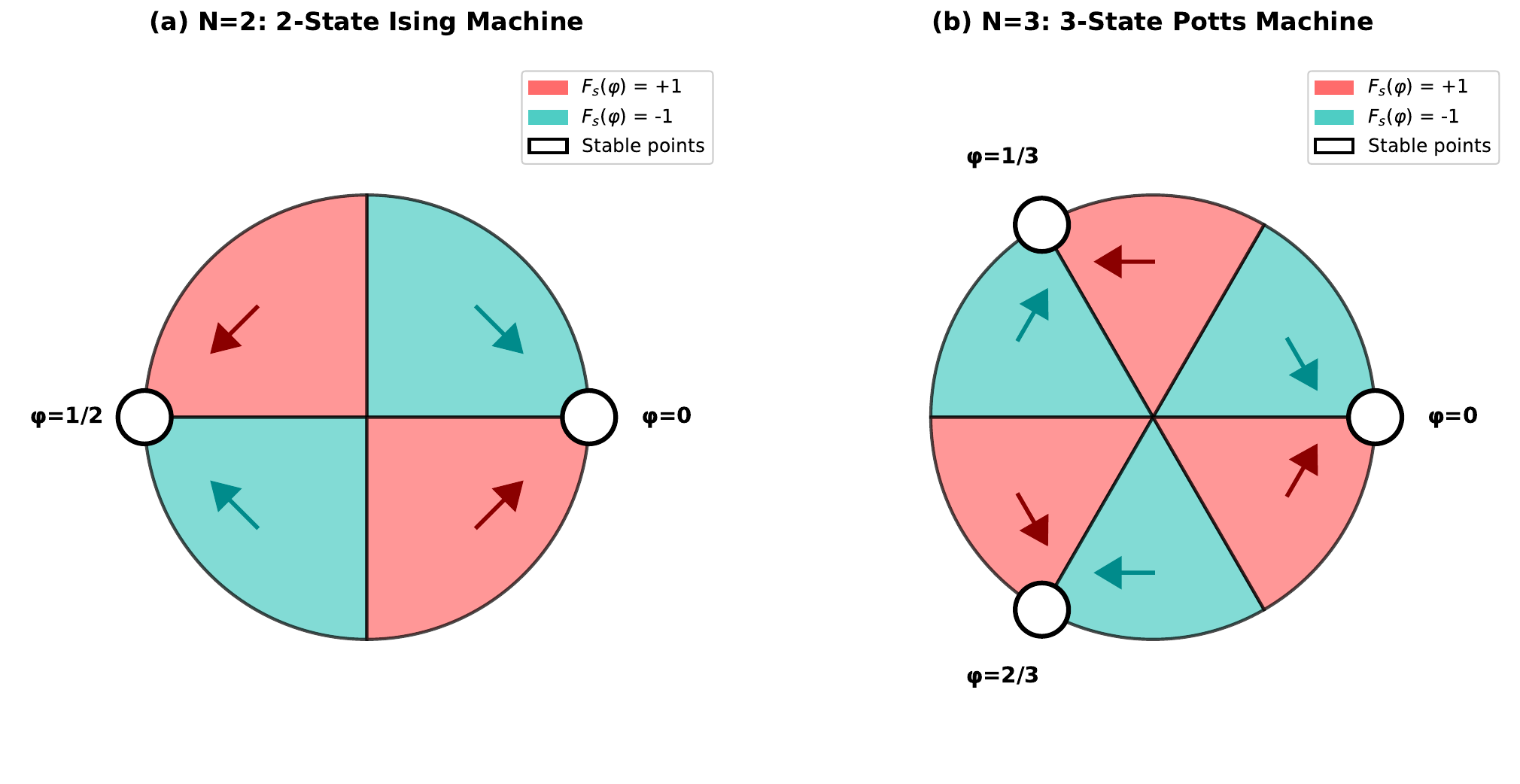}
     \caption{$F_s$ function producing synchronization gradients based on the current phase of the oscillator demonstrated on the unit circle for Ising ($N=2$) and 3-state Potts ($N=3$) configuration}
    \label{fig:unit_circle}
\end{figure}

$F_c(\psi)$ provides positive coupling when oscillators are close in phase and negative coupling when opposite, while $F_s(\phi)$ creates stable regions at $\phi =\{ 0, 0.5\}$ for $N=2$ or $\phi = \{0, 1/3, 2/3\}$ for $N=3$, corresponding to spin states or colors. Figure \ref{fig:unit_circle} illustrates these synchronization dynamics on the unit circle.

The critical advantage of these simplified functions is their computational efficiency in fixed-point arithmetic. When phases are represented in binary fixed-point format with the radix point before the most significant bit (e.g., $\phi = 0.b_1b_2b_3\ldots b_{n}$), the modulo operations are inherently handled by the representation. Evaluating $F_c(\phi_i - \phi_j)$ reduces to checking the most significant bit after subtraction, while $F_s^{(2)}(\phi_i)$ requires checking the second bit. For $F_s^{(3)}$, the top three bits are examined using simple comparators. These operations require no arithmetic, only bit inspection.

\section{Proposed Algorithm and Hardware Architecture}

The phase dynamics are numerically solved in fixed point arithmetic using the Forward Euler method~\cite{forward_euler}, which iteratively updates the phase of each oscillator based on the computed time derivative.
The phase update equation in a Forward Euler time step defined by the  integration time step parameter $h$ is:

\begin{equation}
\label{eq:eq5}
\phi_i \leftarrow \phi_i - h\left[\sum_{j=1}^{n} J_{ij} \cdot F_c(\phi_i - \phi_j) + F_s(\phi_i)\right]
\end{equation}

\subsection{Numerical Algorithm}

Algorithm 1 describes the phase evolution in a single iteration of the forward Euler loop for a single oscillator $i$. The simplified $F_c$ and $F_s$ functions defined in Equations~\eqref{eq:eq2}, \eqref{eq:eq3}, and \eqref{eq:eq4} compute the coupling sum using the phases of the set $\mathcal{N}(i)$ of connecting oscillators. The aggregated coupling sum is then scaled by the timestep parameter $h$ to produce the phase gradient that gets subtracted from the current phase as in Equation~\eqref{eq:eq5}. Synchronization term $F_s$ is introduced after a predetermined settling period during which the system evolves under coupling interactions only, controlled with the $sync\_enable$ signal.

% \begin{algorithm}
% \caption{Phase Update of a Single Oscillator}
% \label{alg:pe_update}
% \begin{algorithmic}[1]
% \State \textbf{Input:} $\phi_i$, $\{\phi_k\}_{k=1}^{q}$, $\{J_{ik}\}_{k=1}^{q}$ $k \in \mathcal{N}(i) $, $h$, $sync\_enable$
% \State \textbf{Output:} $\phi_i$ // Updated phase
% \State $\text{coupling\_sum} \gets 0$
% \For{each neighbor $k \in \mathcal{N}(i)$}
%  // Evaluate coupling function $Fc$ per neighbor and accumulate
%      \State $\text{coupling\_sum} \gets \text{coupling\_sum} + J_{ik} \cdot F_c(\phi_i - \phi_k)$
% \EndFor
% \If{$sync\_enable = 1$}
%      \State $\text{coupling\_sum} \gets \text{coupling\_sum} +F_s(\phi_i)$   // Evaluate synchronization function
% \EndIf
% \State $\phi_i \gets \phi_i - h \cdot (\text{coupling\_sum} + F_s)$
% \end{algorithmic}
% \end{algorithm}

\begin{algorithm}
\caption{Phase Update of a Single Oscillator}
\label{alg:pe_update}
\begin{algorithmic}[1]
\State \textbf{Input:} $\phi_i$, $\{\phi_k\}_{k \in \mathcal{N}(i)}$, $\{J_{ik}\}_{k \in \mathcal{N}(i)}$, $h$, $sync\_enable$
\State \textbf{Output:} $\phi_i$ // Updated phase
\State $\text{coupling\_sum} \gets 0$
\For{each neighbor $k \in \mathcal{N}(i)$}
    \Statex \hspace{\algorithmicindent} // Evaluate coupling function $F_c$ and accumulate
    \State $\text{coupling\_sum} \gets \text{coupling\_sum} + J_{ik} \cdot F_c(\phi_i - \phi_k)$
\EndFor
\If{$sync\_enable = 1$}
    \Statex \hspace{\algorithmicindent} // Evaluate synchronization function $F_s$
    \State $\text{coupling\_sum} \gets \text{coupling\_sum} +F_s(\phi_i)$
\EndIf
\State $\phi_i \gets \phi_i - h \cdot \text{coupling\_sum}$
\end{algorithmic}
\end{algorithm}
\begin{figure}[ht]
    \centering
    \includegraphics[width=\columnwidth]{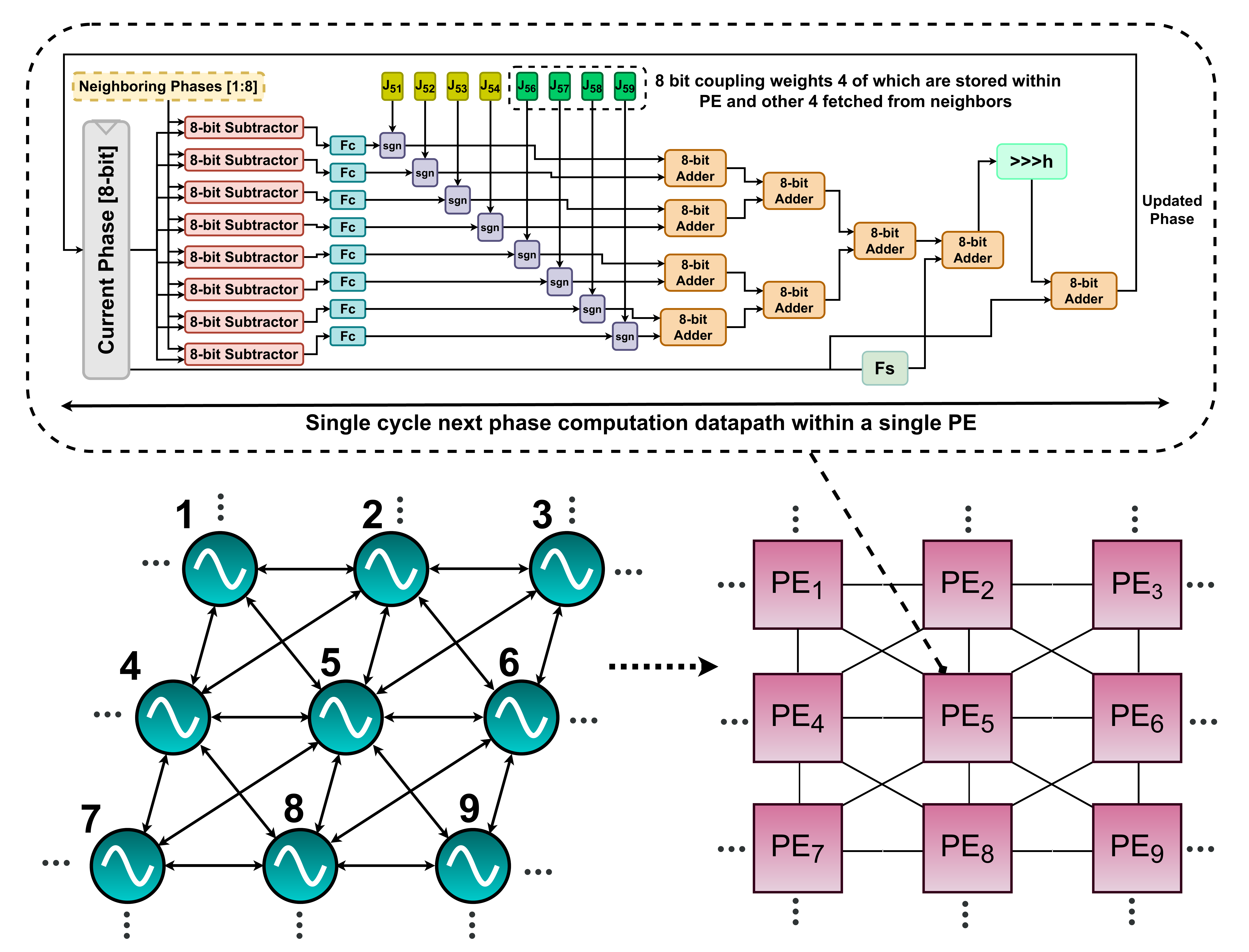}
     \caption{An array of coupled oscillators mapped to an array of PEs and the single cycle phase update computation within a PE}
    \label{fig:array_pe}
\end{figure}

\subsection{Proposed Hardware Architecture}

The proposed hardware architecture implements the OIMs/OPMs as a grid of interconnected processing elements (PEs) in king's graph topology. Each PE represents a single oscillator and executes Algorithm \ref{alg:pe_update} in fixed-point arithmetic to compute phase updates in a single clock cycle per Forward Euler iteration. Figure~\ref{fig:array_pe} shows the PE architecture and the grid of PEs mapping a system of coupled oscillators. Phase information is exchanged through dedicated 8-bit interconnects between PEs. In this topology, the neighbor set $\mathcal{N}(i)$ for each interior PE $i$ consists of the 8 spatially adjacent PEs, with edge and corner PEs on the grid having fewer neighbors.

Each PE maintains its phase state in an 8-bit register, and the coupling weight precision is also selected as 8-bits. A custom bit-accurate simulator is used to evaluate different bit-width configurations, showing that accuracy improves negligibly beyond 8 bits. To prevent duplication of coupling weight registers for each interaction pair, each PE contains four 8-bit coupling weight registers, providing weights to four neighbors while reading weights from the other four neighbors. During each clock cycle (one Forward Euler iteration), all PEs concurrently fetch eight 8-bit phases of neighboring PEs and four 8-bit coupling weights through dedicated interconnect with neighbor PEs, compute the phase gradient according to the simplified Kuramoto dynamics using local arithmetic and logic units as shown in Figure~\ref{fig:array_pe}, and update the local phase $\phi_i$. This fully parallel execution enables a single cycle latency per iteration. 

\begin{figure*}[t] % <-- figure* spans both columns and [t] places it at the top
    \centering

    %--- Left subfigure ---
    \begin{subfigure}{0.32\textwidth}
        \centering
        \includegraphics[width=\textwidth]{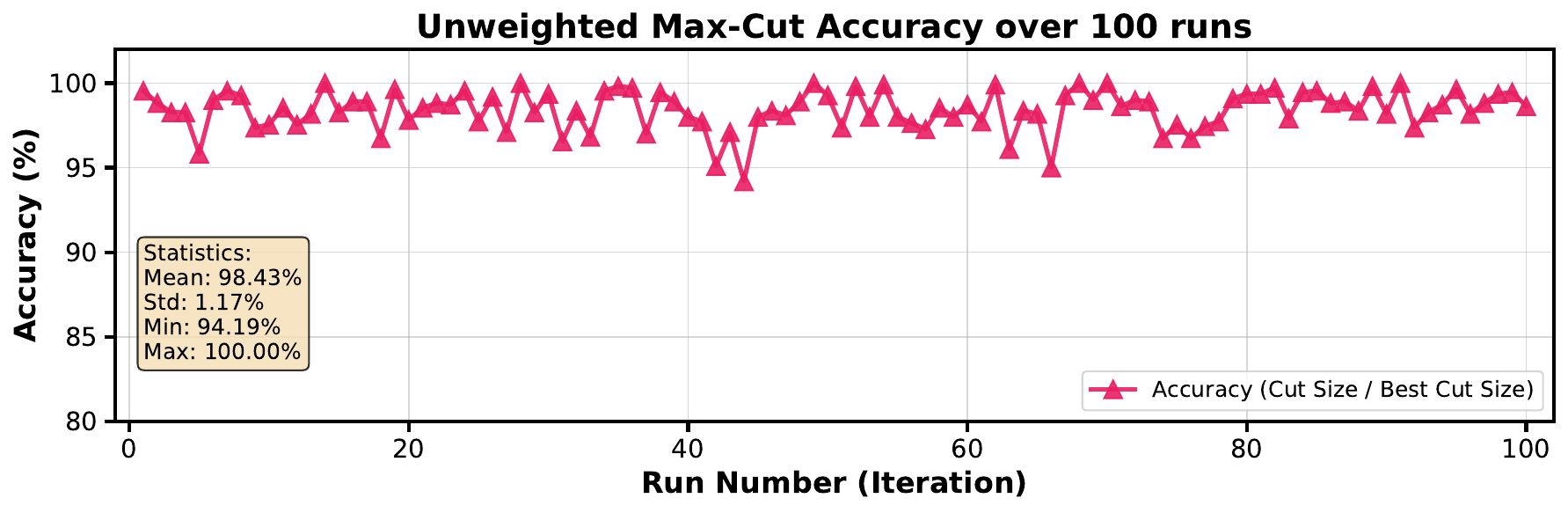}
    \end{subfigure}
    \hfill
    %--- Middle subfigure ---
    \begin{subfigure}{0.32\textwidth}
        \centering
        \includegraphics[width=\textwidth]{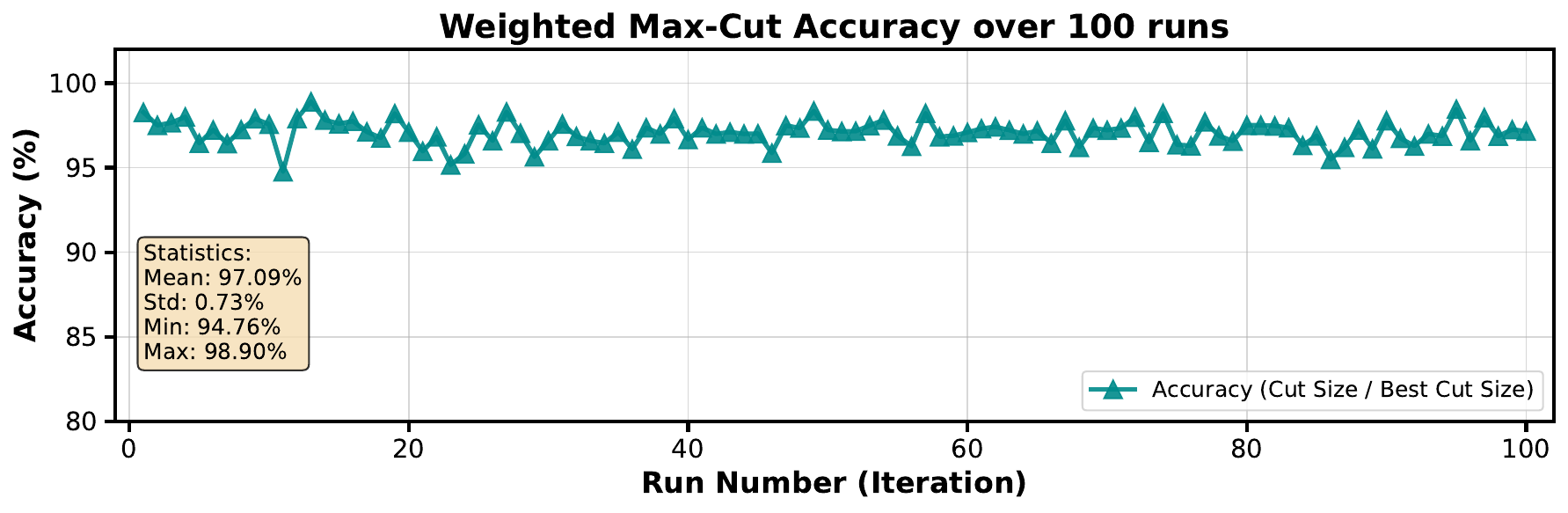}
    \end{subfigure}
    \hfill
    %--- Right subfigure ---
    \begin{subfigure}{0.32\textwidth}
        \centering
        \includegraphics[width=\textwidth]{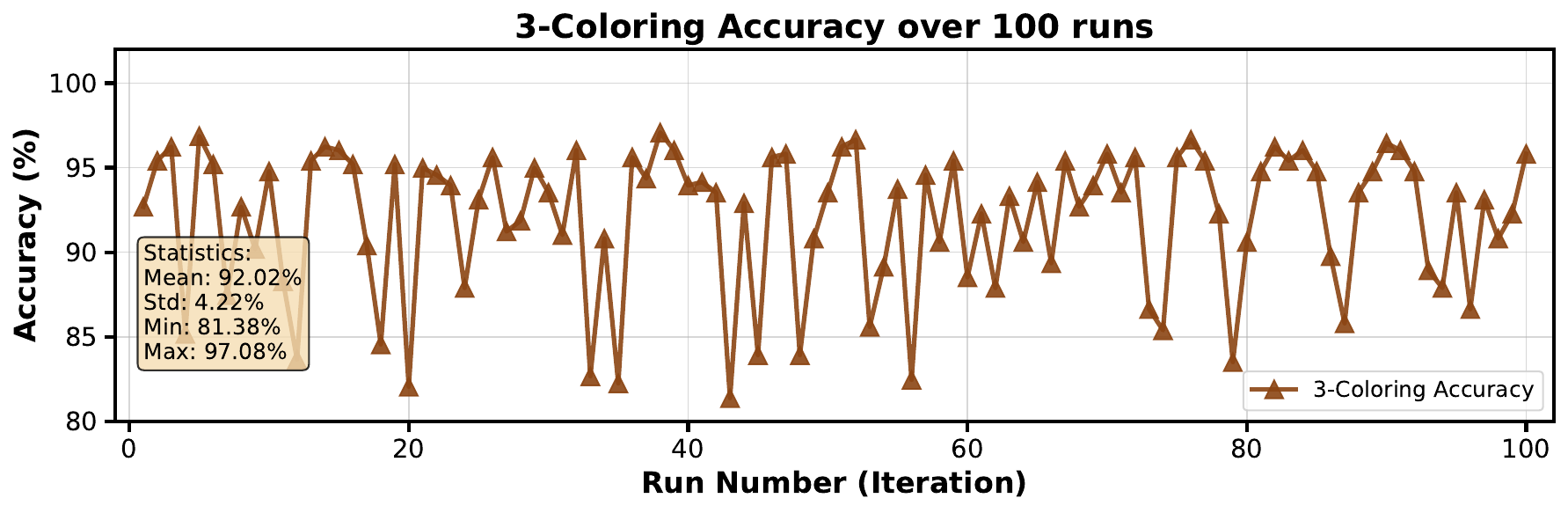}
    \end{subfigure}

    \caption{Accuracy ranges over 100 runs on the hardware of 400 node unweighted, weighted max-cut problems in OIM mode and a 400 node 3-coloring problem in OPM mode, left to right}
    \label{fig:accuracy_plots}
\end{figure*}

The PE microarchitecture consists of 17 8-bit {adder/subtractors for performing the addition and subtraction operations required by Algorithm \ref{alg:pe_update}, along with logic for bit inspection and conditional operations for $F_c$ and $F_s$ computation. These  optimizations reduce the circuit complexity. The step size parameter $h$ is constrained to negative powers of two (e.g., $h = 2^{-6}$), allowing multiplication by $h$ to be implemented as a simple right-shift operation (performed by a barrel shifter) rather than requiring a full multiplier, significantly reducing both area and critical path delay. The coupling term $J_{ij} \cdot F_c$ computation is simplified by 2:1 multiplexing rather than multiplication, where the sign bit of $F_c$ selects either $J_{ij}$ or $-J_{ij}$ for accumulation. This operation is shown by `sgn' blocks in Figure~\ref{fig:array_pe}, where the sign of $J_{ij}$ is determined by 8 multiplexers computing $F_c$ on phase differences in parallel.

% The modular architecture is scalable in multiple dimensions. The number of connected neighbors per PE can be increased beyond eight through additional parallel processing units or reduced through time-multiplexing of the computation, trading area for latency.
%The grid size can be expanded by replicating PEs in larger grid configurations, with the predominantly local connectivity pattern enabling efficient physical routing.

\begin{figure}[H]
    \centering
    \includegraphics[width=0.9\columnwidth]{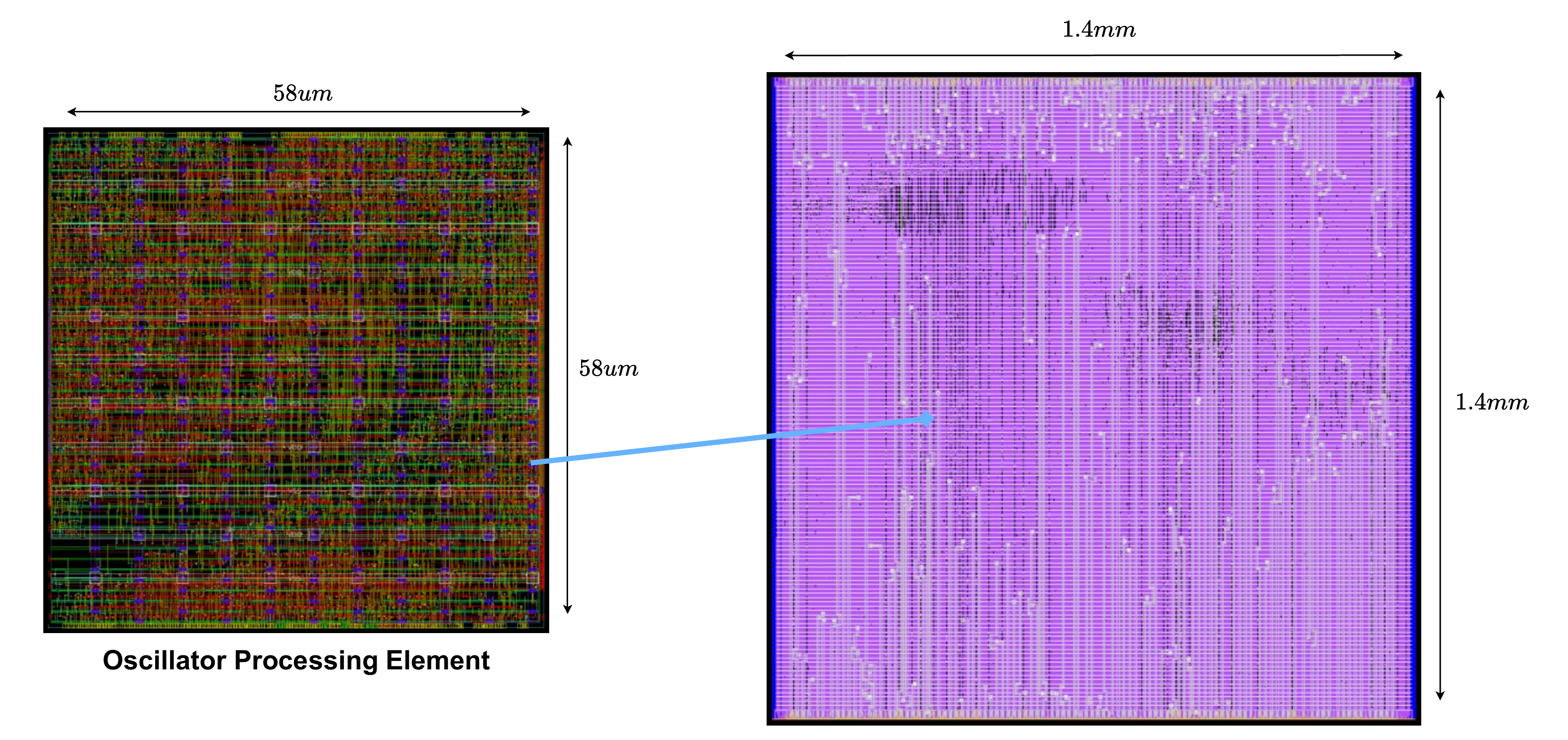}
     \caption{Layout of a single PE (left) and the complete chip architecture (right) containing interconnected PEs}
    \label{fig:layout}
\end{figure}

% \begin{figure}[ht]
%     \centering
%     \includegraphics[width=\columnwidth]{pe_diagram.pdf}
%      \caption{The next phase computation within the single-cycle PE}
%     \label{fig:array_pe}
% \end{figure}

% \begin{figure}[ht]
%     \centering
%     \includegraphics[width=\columnwidth]{array_pe.pdf}
%      \caption{An array of coupled oscillators mapped to an array of PEs}
%     \label{fig:array_pe}
% \end{figure}

\subsection{Physical Implementation}

To evaluate the performance of the proposed architecture, a 20x20 grid of interconnected PEs is implemented as a prototype.
The proposed 20x20 PE grid prototype is implemented using a 65nm CMOS technology. The modular near-neighbor architecture ensures efficient physical routing by utilizing a predominantly local connectivity pattern. Figure \ref{fig:layout} shows the physical layout of both a single processing element and the complete 20×20 array. Each PE occupies approximately 58~$\micro m$~×~58~$\micro m$ (0.0034~$mm^2$), which is  5.4 times larger in area than a ring oscillator unit cell in a similar near-neighbor topology~\cite{chris_kim_nat}. The total core area, including the 400 PEs, interconnect routing, coupling weight registers, control logic, and shift-register interface (for programming the chip), is 1.4~$mm$~×~1.4~$mm$ (1.96~$mm^2$).

\section{Experimental Results}

 The prototype 20x20 array implementation is tested and verified by post-layout simulations to operate at a 200 MHz clock frequency. Power analysis based on post-layout simulations indicates the hardware consumes an estimated 95 mW when all PEs are active (i.e. mapping a problem node). Due to the probabilistic nature of the Ising/Potts model, different runs (i.e. from random initial conditions till convergence) result in different solutions, therefore 100 repeated runs are performed for each problem to explore the solution space.  

\subsection{Unweighted and Weighted Max-Cut Problem Evaluation}

The architecture is evaluated on a custom-generated 400-node unweighted and a 400-node weighted max-cut problem in king's graph topology in the OIM mode. Weighted max-cut problem is generated by randomized fp32 weights quantized into 8 bits. With 1,000 Forward Euler iterations (empirically found to be sufficient for convergence of the selected 400 node problems) per solution, the ASIC chip executes a single run in 5~$\mu s$, able to generate up to 200k different solutions per second. Figure \ref{fig:accuracy_plots} shows accuracy across 100 independent runs on the problems, where accuracy is the ratio of obtained cut size to the best cut (i.e. exact solution). Results demonstrate consistent performance with mean accuracy of 98.43\% and standard deviation of 1.17\%, with minimum accuracy of 94.19\% and 8 optimal (100\% accuracy) solutions achieved over 100 runs.

\subsection{Graph 3-Coloring Evaluation}

A custom generated 3-colorable 400 node graph is mapped onto the 20×20 grid in the OPM mode. The hardware achieves mean coloring accuracy of 92.02\% with standard deviation of 4.22\%, ranging from 81.38\% to 97.08\% across 100 runs. The increased variance and lower mean compared to max-cut result from the 3-bit quantization in $F_s^{(3)}$ where stable points do not align precisely with ideal values of 0, 1/3, and 2/3. Increased number of inspected bits for discretization would increase accuracy at the cost of additional hardware.

% \begin{figure}[htbp]
%     \centering

%     \begin{subfigure}{\columnwidth}
%         \centering
%         \includegraphics[width=\columnwidth]{unweighted_acc.pdf}
%     \end{subfigure}

%     \begin{subfigure}{\columnwidth}
%         \centering
%         \includegraphics[width=\columnwidth]{weighted_acc.pdf}
%     \end{subfigure}

%     \begin{subfigure}{\columnwidth}
%         \centering
%         \includegraphics[width=\columnwidth]{coloring_acc.pdf}
%     \end{subfigure}

%     \caption{Accuracy ranges over 100 runs on the hardware of 400 node unweighted, weighted max-cut problems and a 400 node 3-coloring problem, top-to-bottom}
%     \label{fig:accuracy_plots}
% \end{figure}

\subsection{Comparison With Other Platforms}

Table \ref{tab:energy_comparison} compares the proposed ASIC with an Nvidia A5000 GPU, an Intel Core i7-5870K CPU leveraging 4 parallel threads, both solving the same 400 node problem as the proposed ASIC, and an analog ring oscillator OIM (ROSC OIM) implementation~\cite{prob_fabric} solving a 560 node unweighted max-cut problem. Although GPU and CPU platforms use more advanced process nodes, the ASIC achieves 12,800 $\times$ speedup over the GPU solution. When normalized by clock cycles, the advantage increases substantially given the 200 MHz operation versus GHz-range processors. Power consumption of the proposed ASIC is at 95~mW per solution, consuming 1042x less energy than the GPU implementation. Both GPU and CPU execute the same 8-bit algorithm despite supporting higher precision for fairness.

\begin{table}[htbp]
\centering
\caption{Comparison of different platforms}
\label{tab:energy_comparison}
\setlength{\tabcolsep}{4pt}
\renewcommand{\arraystretch}{1.15}
\resizebox{\columnwidth}{!}{%
\begin{tabular}{@{}lccccc@{}}
\hline
\textbf{Implementation} & \textbf{\shortstack{Run\\ Time}} & \textbf{Technology} & \textbf{Power} & \textbf{Accuracy} & \textbf{\shortstack{Coupling \\ Resolution}} \\
\hline
This Work & 5 $\mu$s & 65nm Digital & 95 mW & 100\% & 8 bits \\
GPU OIM/OPM & 64 ms & GPU (A5000) & 99 W & 100\% & 8+ bits \\
ROSC OIM~\cite{prob_fabric} & 50 ns & 65nm Analog & 23 mW & 92\% & 2 bits \\
%ROSC OPM \cite{iccad_potts} & 11 ns & 65 nm Analog & 155 mW  & 92\% & 1 bit \\
CPU OIM/OPM & 35.4 s & CPU (i7) & N/A & 100\% & 8+ bits \\
\hline
\end{tabular}%
}
\end{table}

The ROSC OIM~\cite{prob_fabric} achieves faster solution time (50~ns), but exhibits lower accuracy (92\% vs 98.4\%) and limited programmability.
ROSC OIMs face significant scalability challenges: achieving 8-bit coupling resolution requires 256 pass transistors (or back-to-back inverters) per connection, and both phases and coupling strengths are analog quantities susceptible to process variations and temperature drift. Large ROSC OIM arrays also suffer from phase drift along the chip, causing phase errors that require correction by post-processing~\cite{chris_kim_nat}. The digitally emulated hardware is free from these analog-related issues.

\subsection{Scalability and Mapping of Other COPs}

The king's graph topology in this work is limited to mapping problems with near-neighbor connectivity with a degree of 8, and up to a maximum of 400 nodes limited by the number of PEs in the ASIC. The proposed algorithm generalizes to arbitrary problem sizes and connectivity patterns. The hardware architecture needs to be updated (left for future work) to support more complex problems such as Gset~\cite{gset} and Satlib~\cite{satlib} benchmarks. Extending the $F_s$ function to account for $N>3$ would enable the N-state Potts model behavior for N-variable COPs, while higher order spin interactions~\cite{higher_order} can be implemented via phase-fixing~\cite{fpim} to create ancillary spins for problems like 3-SAT~\cite{high_quality_oim}. Many other COPs are also accessible through problem reducibility~\cite{reducibility}. 

% The proposed hardware can map other COPs without significant architectural changes. Extending the $F_s$ function to account for $N>3$ would allow N-state Potts model behavior to solve N-variable COPs. Higher order spin interactions~\cite{higher_order} can also be implemented by phase-fixing~\cite{fpim} to create ancillary spins solving problems like 3-SAT~\cite{high_quality_oim} natively. Many other COPs can be mapped to the existing architecture via the reducibility of COPs to each other~\cite{reducibility}.

% The proposed algorithm generalizes to arbitrary problem size and connectivity patterns beyond king's graph. The architecture can be extended to include a more complex and configurable data movement network, and time-sliced scheduling of computation to enable mapping of larger, denser, and more irregular problems such as the Gset benchmarks~\cite{gset}. 

\section{Conclusion}

This work presents a custom ASIC architecture for digitally emulating oscillator-based Ising machines in fixed-point arithmetic. A 20×20 PE grid of king's graph topology in 65nm CMOS achieves 5~$\micro s$ solution time with 95~$mW$ power consumption while reaching 97-100\% max accuracy on 400-node max-cut and coloring problems. These results demonstrate the viability of an algorithmically co-designed digital ASIC for oscillator-based Ising and Potts machines.

% \section {Acknowledgements}
% This material is based upon work supported by the National Science Foundation under Grants No. 1409014, 1816857.

\bibliographystyle{unsrt}
\bibliography{ref}

\end{document}